\documentclass[preprint2, letterpaper]{aastex}
\usepackage{amsmath}
\slugcomment{Draft 1: Done with abstract and conclusions}
\shorttitle{EoR imaging:effect of foregrounds and imaging algorithms}
\shortauthors{Vedantham et al.}
\begin{document}
%

\title{Imaging the Epoch of Reionization: limitations from foreground confusion and imaging algorithms}
\author{Harish Vedantham, N. Udaya Shankar, and Ravi Subrahmanyan}
\affil{Raman Research Institute, Bangalore, India}
\email{harish@rri.res.in}

\begin{abstract}
Tomography of redshifted 21 cm transition from neutral Hydrogen using Fourier synthesis telescopes is a promising tool to study the Epoch of Reionization (EoR). Limiting the confusion from Galactic and Extragalactic foregrounds is critical to the success of these telescopes. Instrumental response or the Point Spread Function (PSF) of such telescopes is inherently 3 dimensional with frequency mapping to the Line of Sight (LOS) distance. EoR signals will necessarily have to be detected in data where continuum confusion persists; therefore, it is important that the PSF has acceptable frequency structure so that the residual foreground does not confuse the EoR signature. This paper aims to understand the 3 dimensional PSF and foreground contamination in the same framework. We develop a formalism to estimate the foreground contamination along frequency, or equivalently LOS dimension, and establish a relationship between foreground contamination in the image plane and visibility weights on the Fourier plane. We identify two dominant sources of LOS foreground contamination---`PSF contamination' and `gridding contamination'. We show that `PSF contamination' is localized in LOS wavenumber space, beyond which there potentially exists an `EoR window' with negligible foreground contamination where we may focus our efforts to detect EoR. `PSF contamination' in this window may be substantially reduced by judicious choice of a frequency window function. Gridding and imaging algorithms create additional `gridding contamination' and we propose a new imaging algorithm using the Chirp Z Transform (CZT) that significantly reduces this contamination. Finally, we demonstrate the analytical relationships and the merit of the new imaging algorithm for the case of imaging with the Murchison Widefield Array (MWA).
\end{abstract}
\section{Introduction}
Several synthesis telescopes such as MWA\footnote{See \url{http://www.mwatelescope.org}}, LOFAR\footnote{See \url{http://www.lofar.org}}, and PAPER\footnote{See \url{http://astro.berkeley.edu/~dbacker/eor}} are currently being built to study the Epoch of Reionization (EoR) by observing the redshifted $21$ cm line transition from neutral hydrogen. While future instruments may map the $21$ cm brightness in three dimensions (frequency corresponds to line of sight), current experiments lack the required sensitivity and instead hope to estimate the power spectrum of 21cm brightness fluctuations. The brightness of these fluctuations is about 5 orders of magnitude smaller than the Galactic and Extragalactic foregrounds. Moreover, removing the effects of the telescope point spread function (PSF) from the synthesis images may not be possible to the level of sensitivity demanded by EoR power spectrum measurements. Consequently, foreground contamination and instrumental effects present a major hurdle to $21$ cm tomography. \\

A host of simulations have demonstrated the potential in current instruments to recover the EoR power spectrum despite thermal uncertainties, foreground contaminants, and instrumental effects. Among instrumental effects, simulations have shown the frequency dependence of PSF to be of particular concern in EoR power spectrum estimation \citep{judd09}. In $21$ cm tomography, frequency maps to the line of sight, and any instrumental variation in frequency can potentially `mimic' line of sight $21$ cm brightness fluctuations. Most of the proposed foreground subtraction algorithms rely on the ability to distinguish smooth spectral features of Galactic and Extragalactic contaminants from the relatively rapid spectral fluctuations (expected from theory) of the $21$ cm EoR signal. Spectral features in the telescope PSF may affect our ability to `fit' and remove undesirable foregrounds. The sidelobe response of confusing foreground contaminants may vary with frequency and generate an instrumental spectral structure at each image pixel---a phenomenon that has been dubbed `mode-mixing' in literature \citep{judd09}. Apart from mode-mixing, calibration errors and choice of gridding algorithms can also lead to undesirable instrumental response in frequency domain.\\

While simulations have evaluated such mode-mixing effects for realistic arrays like MWA \citep{liu09,judd09}, this paper presents an approach that is analytical, and hence, more general. In particular, the relationship between the PSF in image coordinates ($lmf$) and the telescope sampling function in Fourier coordinates ($uvf$) is established, and the mode-mixing problem is cast in Fourier space where the telescope measurements are really made. In this formalism, implications for choice of experimental parameters and foreground subtraction algorithms are discussed. Apart from the analytical treatment, various practical data processing operations that influence the PSF are discussed using an example of snapshot imaging with MWA.\\

The remainder of the paper is organized as follows. Section \ref{sec:analytical} defines the problem analytically and derives an expression relating the line of sight foreground contamination to the weighting function in Fourier space. To better understand the relationships established in Section \ref{sec:analytical}, Section \ref{sec:fully_filled} applies them to the simple case of imaging a point source with a complete visibility distribution. The analytical treatment in Section \ref{sec:analytical} does not take into account the effects of discretization of visibilities in the re-gridding and imaging algorithms. Section \ref{sec:gridding_algo} discusses these effects and complements the results of Section \ref{sec:analytical}. In particular, Section \ref{sec:gridding_algo} describes a new imaging algorithm based on the Chirp Z Transform (CZT) that alleviates the problem of foreground contamination. Section \ref{sec:mwa_example} demonstrates the concepts developed in Sections \ref{sec:analytical}, \ref{sec:fully_filled}, and \ref{sec:gridding_algo} through simulating a snapshot observation with MWA. Section \ref{sec:conclusions} draws up conclusions and proposals for future work. 

\section{Analytical treatment}
\label{sec:analytical}
We first analytically evaluate the effect of a frequency variant PSF at a given image pixel due to an unsubtracted point source by evaluating its contribution to foreground contamination in LOS wavenumber space.  We call this contamination `PSF contamination'. A single point source is considered here to attain an intuitive understanding of `PSF contamination'. Generalization for a realistic sky will be made in subsequent sections. Array element locations are assumed to be coplanar.
\subsection{Notation} 
We define the co-ordinate system to avoid ambiguity. The baselines are defined on a local plane whose origin is at the latitude and longitude of the array center. A two dimensional vector ${\bf x}=[x\,\,y]$ (in meter units) represents the projected baseline on the local plane. The corresponding baseline ${\bf u}=[u\,\,v]$ (in wavelengths) at a frequency $f$ is then given by ${\bf u} = {\bf x}.(f/c)$, where $c$ is the speed of light. A two-dimensional direction vector ${\bf l}$ towards any direction is represented by its direction cosines: ${\bf l} = [l\,\,m]$. An interferometer observation gives a visibility cube with two spatial axes representing the baseline and a frequency axis. An image cube on the other hand has two axes representing direction cosines on the sky and a third frequency axis. Assuming co-planarity of array elements, within the framework of Fourier synthesis imaging, ${\bf u}$ and ${\bf l}$ are Fourier conjugates.
\subsection{Weighting function}
\label{subsec:weighting_function}
The telescope sampling function is represented as $\bar{W}({\bf x}, f)$, which is the weight assigned to the baseline represented by the vector $\bf{x}$, and includes any visibility taper applied prior to imaging. In general, $\bar{W}({\bf x})$ depends on instrumental, observational and processing parameters. Among instrumental parameters is the antenna layout that determines the set of available baselines. Observational parameters, which include pointing direction, duration of rotation synthesis, and mode of observation---drift scan or track mode, define the $uv$ coverage. Observational parameters along with the antenna layout determines the `natural' weight at each ${\bf x}$---number of visibilities falling into the pixel centered at ${\bf x}$. Among processing parameters are the visibility grid size and extent, convolution kernel used to grid the visibilities, and any visibility mask and/or taper used during imaging. The weighting function may be factored into a frequency independent geometric component, a frequency dependent `taper' component, and a frequency window function:
\begin{equation}
\bar{W}({\bf x}, f)=W({\bf x})\,T({\bf x},f)\,B(f).
\end{equation}
Instrumental and observational parameters are geometric in nature and hence frequency independent. They may be sufficiently represented by $W({\bf x})$. Processing parameters like visibility `taper' function and visibility masks are frequency dependent and are represented by $T({\bf x},f)$. $B(f)$ is the window function in frequency assigned to each visibility.
\subsection{Derivation}
\label{subsec:derivation}
An image made with a weighted baseline distribution $\bar{W}({\bf x}, f)$ has a frequency dependent PSF $\tilde{A}\left({\bf l}, f\right)$, where ${\bf l}$ is the two dimensional unit vector representing direction on the sky. Tilde notation is used for image domain functions as opposed to their Fourier domain counterparts. Consider the sidelobe contribution at zenith (${\bf l}=[0\,\,\,0]$) from a source with flux density of unity and spectral index of zero located at any ${\bf l}$. This contribution $\tilde{S}({\bf l},f)$ varies with frequency due to a changing PSF. We wish to compute the energy in these spectral variations on different scales in terms of the telescope weighting function $\bar{W}({\bf x}, f)$. \\

PSF $\tilde{A}({\bf l},f)$ is the Fourier Transform of the $uv$-weighting function $A({\bf u},f)$. We may represent this as
\begin{equation}
\tilde{A}({\bf l},f)=\frac{\int_{-\infty}^{\infty}du\int_{-\infty}^{\infty}dv\, A({\bf u},f)  \exp(-i2\pi {\bf u}.{\bf l})}{\int_{-\infty}^{\infty}du\int_{-\infty}^{\infty}dv\, A({\bf u},f)},
\end{equation}
where the denominator is included to obtain $\tilde{A}({\bf l},f)$ as the normalized PSF for all frequencies, and essentially does bandpass calibration of visibilities. The inclusion of $\tilde{A}({\bf l},f)$ essentially does bandpass calibration of visibilities and results in a PSF that is normalised at all frequencies. Any additional frequency weighting may be absorbed into the frequency window function $B(f)$. The function $A$ is just a scaled version of the function $\bar{W}$ through the scaling relationship given by ${\bf u}={\bf x}.(f/c)$. We may represent this scaling relationship as
\begin{equation}
A({\bf u},f) = \bar{W}\left({\bf u}\frac{c}{f},f\right) = \bar{W}\left({\bf x},f\right).
\end{equation}
The quantity $uc/f$ has units of distance and is frequency invariant: it is simply the baseline distance in meters. We use this to make the substitution $x=uc/f$, $y=vc/f$ and get
\begin{eqnarray}
 \tilde{A}({\bf l},f) & = & \frac{1}{\mathcal{A_W}}\int_{-\infty}^{\infty} dx \int_{-\infty}^{\infty} dy \nonumber \\ 
 & & \bar{W}({\bf x},f) \exp\left(-i2\pi {\bf x}.{\bf l}\frac{f}{c}\right),
\label{eqn:beam}
\end{eqnarray}
%
where $\mathcal{A_W}=\int_{-\infty}^{\infty} dx \int_{-\infty}^{\infty} dy \, \bar{W}({\bf x},f)$ is the area under $\bar{W}$. The importance of this substitution is that it expresses all baselines in units of meter as opposed to wavelength and makes them independent of frequency.\\

Consider a point source with a flux density of unity and a spectral index of zero located at ${\bf l}$. The response at zenith due to a sidelobe on this source is given by
\begin{equation}
\tilde{S}({\bf l},f) = \tilde{A}({\bf l},f)
\end{equation}
We are interested in the energy in $\tilde{S}({\bf l},f)$ at different frequency scales. In other words, to evaluate `PSF contamination', we have to compute the power spectrum $\tilde{S}({\bf l},\eta)$ by Fourier transforming $\tilde{S}({\bf l},f)$ with frequency offset $\Delta f$ and time $\eta$ as Fourier conjugates. We use $\Delta f$ instead of $f$ because the spatial EoR fluctuations are evaluated about an origin $r_0$ corresponding to a frequency $f_0$ and redshift $z_0$.\\

$\tilde{S}(\eta,{\bf l})$ may be expressed as
\begin{equation}
\tilde{S}({\bf l},\eta) = \int_{-\infty}^{\infty}d(\Delta f)\, B(f) \tilde{A}({\bf l},f) \exp(-i2\pi\eta\Delta f).
\end{equation}
Using $\Delta f=f-f_0, \,\,f_0=(f_L+f_H)/2$, where $f_L$ and $f_H$ are the lowest and highest frequencies in the instrument bandpass, and writing the conjugate variable as $f$ instead of $\Delta f$ we get
\begin{eqnarray}
\tilde{S}({\bf l},\eta) & = & \exp(i2\pi\eta f_0)\int_{-\infty}^{\infty} df\nonumber \\
 & & B(f) \tilde{A}({\bf l},f) \exp(-i2\pi\eta f).
\end{eqnarray}
Substituting from Equation \ref{eqn:beam} for $\tilde{A}({\bf l},f)$ and interchanging the order of integrations we get
\begin{multline}
\tilde{S}({\bf l},\eta) = \exp(i2\pi\eta f_0) \int_{-\infty}^{\infty}dx\int_{-\infty}^{\infty}dy\, W({\bf x}) \\
\int_{-\infty}^{\infty}df\, \frac{T({\bf x},f)}{\mathcal{A_W}}B(f) \exp\left(i2\pi f\left(\frac{{\bf x.l}}{c}-\eta\right)\right).
\label{eqn:general_expr}
\end{multline}
The inner integral, in general, may have to be evaluated numerically. However, to have an insight into the nature of $\tilde{S}({\bf l},\eta)$, we evaluate Equation \ref{eqn:general_expr} for the case where there is no taper or mask applied in the $uv$ domain. In this case, $T({\bf x},f) \equiv 1$, and $\mathcal{A_W}$ is frequency independent. The inner integral is now just a scaled and shifted version of the Fourier transform of the frequency window function. Consider a rectangular frequency window defined by
\begin{equation}
B(f) = \left\{ \begin{array}{rl}
         \frac{1}{\mathcal{B}} & \mbox{ if $f\in [f_L,\,f_H]$} \\
         0 &\mbox{ otherwise,}
       \end{array} \right.
\label{eqn:rect_bandpass}
\end{equation}
where $\mathcal{B}=f_H-f_L$ is the frequency bandwidth.
The inner integral now evaluates to
\begin{eqnarray}
\tilde{B}\left(\frac{{\bf x.l}}{c}-\eta\right) & = & \frac{\sin\left(\pi\mathcal{B}\left(\frac{{\bf x.l}}{c}-\eta\right)\right)}{\pi\mathcal{B}\left(\frac{{\bf x.l}}{c}-\eta\right)}\nonumber \\
 & &  \exp\left[2\pi f_o\left(\frac{{\bf x.l}}{c}-\eta\right)\right].
\label{eqn:bw_sinc}
\end{eqnarray}
This gives us the relationship we are looking for:
\begin{eqnarray}
\tilde{S}({\bf l},\eta) & = & \frac{\exp(i2\pi\eta f_o)}{\mathcal{A_W}}\int_{-\infty}^{\infty}dx\int_{-\infty}^{\infty}dy \nonumber \\
 & & W({\bf x})\, \tilde{B}\left(\frac{{\bf x.l}}{c}-\eta\right).
\label{eqn:final_relationship}
\end{eqnarray}
The power spectrum $\tilde{S}({\bf l},\eta)$ is a convolution of the weighting function $W({\bf x})$ and the convolving kernel $\tilde{B}$, which is a shifted version of the Fourier transform of the frequency window function $B(f)$.
\section{Ideal case of complete visibility coverage}
\label{sec:fully_filled}
\begin{figure}[h]
\epsscale{1.0}
\plotone{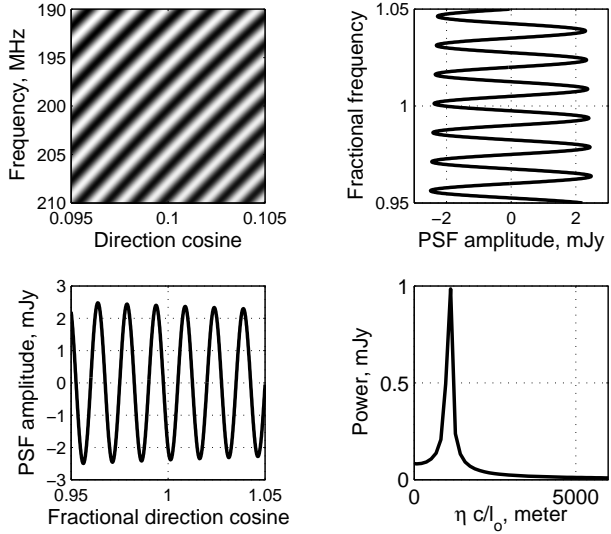}
\caption{Plots showing an example of PSF sidelobe structure in frequency for a uniformly weighted visibility distribution with complete coverage ($W(x)=1$ for $|x|<1000$, $W(x)=0$ otherwise). Top left panel shows a small section of the PSF in the vicinity of $(l_o,\,f_o)=(0.1,\,200$ MHz$)$. Top right panel shows PSF variation at $l_o$ as a function of fractional frequency, $\delta f/f_o$. Bottom left panel shows PSF variation at $f_o$ as a function of fractional directional cosine, $\delta l/l_o$. Bottom right panel shows the PSF Fourier transformed with respect to frequency to reveal the `PSF contamination' in $\eta$ space due to a point source at $l_o$.\label{fig:lf_invariance_unif}}
\end{figure}  
This section provides an insight into the terms and equations of the analytical derivation, by understanding them in the case of a visibility distribution with complete coverage. Complete coverage refers to the case where there exists at least one visibility measurement in every pixel in the $uv$ grid at all frequencies within the instrument bandwidth, and over the spatial frequency range spanned by the array configuration. A key idea in the analytical treatment of Section \ref{sec:analytical} is the invariance of the PSF in ${\bf l}f$. This concept is manifested in the invariance of Equation \ref{eqn:beam} in ${\bf l}f$ for a frequency invariant weighting. Frequency invariant weighting implies a `taper' function $T(x)$ that is invariant in frequency and dependent only on $x$. Figure \ref{fig:lf_invariance_unif} shows an image of the PSF in $lf$ domain for a uniformly weighted visibility distribution with complete coverage. The plot extends over a $10\%$ spread in both frequency and direction cosine ($\Delta l/l_o=\Delta f/f_o=0.1$). Due to the $lf$ invariance, we expect the PSF to have the same value at all points defined by $lf=l_of_o$. Consequently, the variation in an interval $\delta l$ of the PSF evaluated at $f_o$, is the same as the variation in an interval $f_o$ of the PSF evaluated at $l_o$, so long as $\delta l/l = -\delta f/f$ or approximately $\delta l/l_o = -\delta f/f_o$. This can be obtained by differentiating $lf=l_of_o$ and then approximating $f$ to $f_o$ and $l$ to $l_o$, which hold for $\delta l/l << 1$ and  $\delta f/f << 1$. Figure \ref{fig:lf_invariance_unif} also evaluates `PSF contamination' in the zenith pixel in $\eta$ space due to PSF sidelobes on a $1$ Jy point source at $l_o$. The localization of this contamination in light of Equations \ref{eqn:bw_sinc} and \ref{eqn:final_relationship} is pivotal to our understanding of `PSF contamination' and deserves a detailed explanation.\\
\begin{figure}[h]
\epsscale{1.0}
\plotone{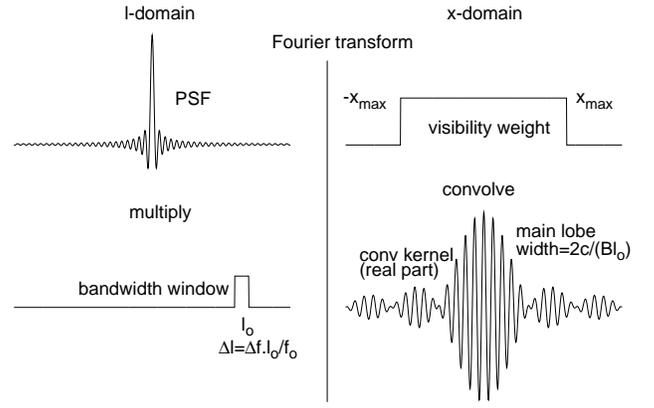}
\caption{Plot depicting the operation in Equations \ref{eqn:bw_sinc} and \ref{eqn:final_relationship}. See Section \ref{subsec:localization} for a detailed explanation of the operations involved.\label{fig:bwsinc_finalexpr}}
\end{figure}  

\subsection{Localization of `PSF contamination'}
\label{subsec:localization}
\begin{figure}[h]
\epsscale{1.0}
\plotone{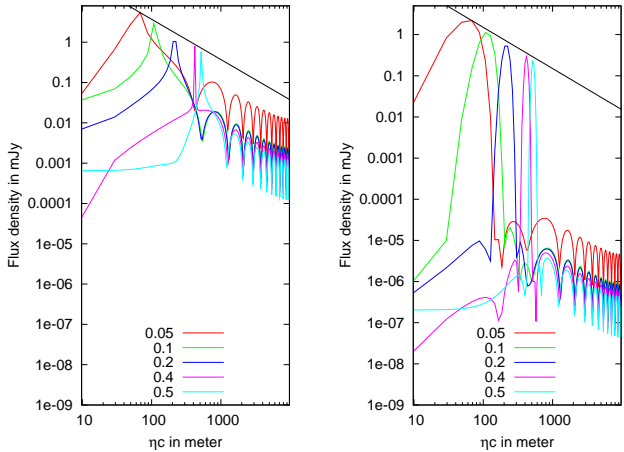}
\caption{`PSF contamination' in $\eta$ space due to $1$ Jy point sources at $l_o=0.05,0.1,0.2,0.4,$ and $0.5$. The left panel shows the contamination for $W(x)$ with uniform weight and a rectangular frequency bandpass, and the right panel shows the contamination for a Blackman-Nuttall window function. For a visibility distribution with complete coverage, the PSF sidelobes are a result of the `top hat' shape of the weighting function $W(x)$, and the convolution in Equation \ref{eqn:final_relationship} has a delta-function like impulse located at $\eta=x_{max}l_o/c$. The magnitudes of such impulses are equal to the sidelobe level at the value of $l_o$ where they are evaluated.  The solid black line is an approximate $1/\eta$ fit to the peaks of the delta-function like impulses. For $\eta > x_{max}l/c$, the sidelobes of the convolving kernel give a response that further falls off as $1/\eta$. The Blackman-Nuttall window substantially lowers contamination in the `EoR window' ($\eta>x_{max}l_o/c$). \label{fig:rel_sll}}
\end{figure}

Consider a single point source at direction cosine $l_o$ entering the zenith pixel through the PSF sidelobe at $l_o$. This is depicted in Figure \ref{fig:bwsinc_finalexpr}. Due to the ${\bf l}f$ invariance, the frequency structure of the PSF at the zenith pixel is same as the sidelobe structure of the PSF in $l$ evaluated at $f_o$. The frequency structure of the PSF in a bandwidth $\delta f$ is identical to the sidelobe structure of the PSF (evaluated at $f_o$) in the interval $[l_o-\delta f\,l_o/(2f_o)\,\,,\,\,l_o+\delta f\,l_o/(2f_o)]$. This truncated version of the PSF can be extracted by multiplying the PSF with a window centered at $l_o$ and having width $\delta f\,l_o/f_o$. A Fourier transform of the product of these two functions is the convolution of their individual Fourier transforms which are the baseline weighting function $W(x)$ and the convolution kernel from Equation \ref{eqn:final_relationship}. It may be seen from Equation \ref{eqn:bw_sinc} that the sinc-function envelope of the convolution kernel has a main lobe width equal to $2c/(l_o\Delta f)$, which in general is far narrower than the visibility range $2x_{max}$. The rapid variation in the convolution kernel is a complex sinusoid in $x$ with frequency $l_o\,f_o/c$. At any $\eta$, the convolution in Equation \ref{eqn:final_relationship} is sensitive to a small range in $W(x)$ that is centered at $x=c\eta/l_o$ with a width $c/(l_o\Delta f)$ which is the main-lobe width of the convolving kernel. For a visibility distribution with uniform weight and complete coverage like the one discussed, owing to the rapid fluctuations of the complex sinusoid, there exists significant response only around $x_{max}$. This explains the delta-function like `PSF contamination' in $\eta$ centered at $\eta=x_{max}\,l_o/c$ (see Figure \ref{fig:lf_invariance_unif}). Sources from different directions in the sky have different values of $l_o$ and hence contribute to foreground contamination at different fairly localized values of $\eta$. The relative magnitude of these impulses is determined by the relative sidelobe levels at the respective directions. Figure \ref{fig:rel_sll} shows the delta-function like `PSF contamination' in $\eta$ for point sources entering the imaged pixel at zenith through different parts of the PSF (different $l_o$). A rectangular window has been assumed in frequency domain. The finiteness of $W(x)$ gives a $1/l$ roll-off in sidelobe levels which corresponds to a $1/\eta$ roll-off in the contamination at different values of $\eta$.
\subsection{Sidelobes of the convolving kernel}
\label{subsec:sidelobes}
We have used the phrase `delta-function like' to describe the `PSF contamination' for a source entering the zenith pixel from the direction $l_o$. The exact profile is described by the sinc envelope of the convolving kernel. Owing to this, the $\eta$ space may be divided into two regions. For $l_o \in [0\,1]$, the delta-function like peaks are confined to $\eta \in [0\,\,x_{max}/c]$. This defines a region of relatively high foreground contamination, say $\mathcal{R}_1$. However, for $\eta > x_{max}/c$, there still exists `PSF contamination' due to the sidelobes of the convolving kernel. The contamination in this regime, say $\mathcal{R}_2$, falls off as $1/\eta$ since the sidelobes of a sinc-function envelope of the convolving kernel have a $1/x$ roll off. While EoR power spectrum estimation for $\eta \in [0\,\,x_{max}/c]$ will suffer from stronger contamination, the level of contamination for $\eta > x_{max}/c$ may be reduced by a judicious choice of a bandpass window $B(f)$ that gives low sidelobe levels in $\mathcal{R}_2$, and hence, $\mathcal{R}_2$ defines an `EoR window' where we may focus our efforts to detect EoR.\\

In essence, a Fourier synthesis array gives a PSF that varies with frequency. If the baseline extent of the array is $x_{max}$, the spectrum of frequency variations will have significant energy only in $\mathcal{R}_1$ defined by $\eta \in [0\,\,,\,\,x_{max}/c]$. The instrument will sample the $\eta$-space up to $\eta_{max}=1/{\mathcal B}$. In ${\mathcal R}_2$ defined by $\eta \in [x_{max}/c\,\,,\,\,1/{\mathcal B}]$ the foreground contamination is due to the $1/\eta$-form sidelobes of the contaminants in ${\mathcal R}_1$. These $1/\eta$ sidelobes are the result of a `top hat' frequency window. A smooth window function in frequency like, for example, a Blackman-Nuttall window \citep{nut81} will substantially reduce the sidelobe contamination in ${\mathcal R}_2$ (by more than three orders of magnitude) at the cost of reduced resolution in $\eta$ space (see Figure \ref{fig:rel_sll}). 
\subsection{Instrumental $k$ space}
\label{subsec:kspace}
\begin{figure}[h]
\epsscale{1.0}
\plotone{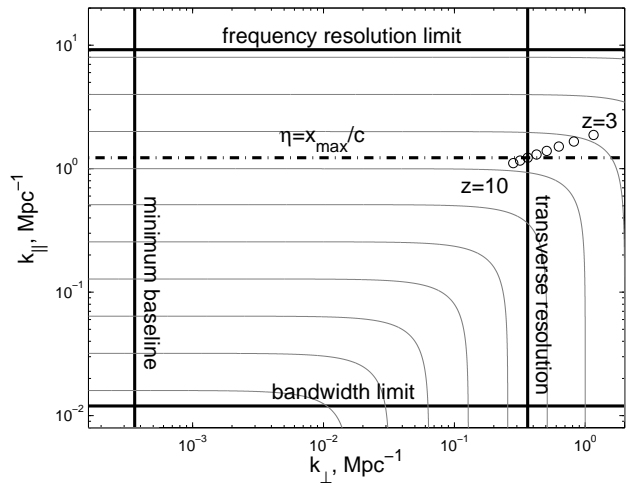}
\caption{Figure showing instrumental response in $k$ space at redshift of $z=8$. The MWA instrument parameters have been used for illustration. A bandwidth limit of $30$ MHz defines a minimum $k_{|| min}$ of about $0.012$ Mpc$^{-1}$, a frequency resolution of $40$ kHz defines a $k_{|| max}$ of about $9.1$ Mpc$^{-1}$, an array extent of $1$ km defines a $k_{\perp max}$ of about $0.3634$ Mpc$^{-1}$, and a minimum baseline of $7.5$ meters gives a $k_{\perp min}$ of about $3.5$ x $10^{-4}$. The region of high foreground contamination $\mathcal{R}_1$ extends up to $\eta=x_{max}/c=3.33\,\mu$sec that corresponds to $k_{||m}$ of about $1.2$ Mpc$^{-1}$. The region between $k_{||m}$ and $k_{|| max}$ is the `EoR window' $\mathcal{R}_2$ and is a region of low foreground contamination. The unfilled black circles define the point $(k_{\perp max},\,\,k_{||m})$ for redshifts of $3,4,5,6,7,8,9,$ and $10$. Each concentric curve is the local of constant $k=\sqrt{k_{\perp}^2+k_{||}^2}$.\label{fig:kspace}}
\end{figure}  
The instrumental $k$ space is an effective space to represent regions and extent of foreground contamination. The instrumental $k$ space is 2 dimensional with the comoving line of sight wavenumber $k_{||}$ and comoving transverse wavenumber $k_{\perp}$ as axes. If $k_x$, $k_y$, and $k_z$ are comoving wavenumbers along three spatial axes with origin at some point about which the EoR fluctuations will be estimated, then \citet{morales04} have shown that 
\begin{multline}
k_{\perp} = \sqrt{k_x^2+k_y^2},\,\,\,\ k_x = \frac{2\pi u}{D_M(z)},\\
  k_y =  \frac{2\pi v}{D_M(z)},\,\,\,\, k_z = k_{||}\approx \eta\frac{2\pi H_of_{HI}E(z)}{c(1+z)^2},
\label{eqn:scales}
\end{multline}
where the approximation holds for a small bandwidth (small redshift range). Here $f_{HI}=1420$ MHz is the rest frequency of $21$ cm HI line emission, $H_0=70$ km sec$^{-1}$ Mpc$^{-1}$ is the present value of the Hubble constant, $E(z)=\left(\Omega_M(1+z_o)^3+\Omega_k(1+z_o)^2+\Omega_{\Lambda}\right)^\frac{1}{2}$, and $D_M(z)$ is the transverse comoving distance at redshift $z$. Figure \ref{fig:kspace} shows a plot of the measurement space (in $k_{\perp}-k_{||}$ domain) for a Fourier synthesis array with complete visibility coverage. A Fourier synthesis array is sensitive to measurements in a region in the $k$ space bounded by some maximum and minimum values of $k_{\perp}$ and $k_{||}$. The highest $k_{\perp}$ mode sampled by the instrument is related to the $uv$ extent of the array and is given by
\begin{equation}
k_{\perp max}=\frac{2\pi x_{max}f_{HI}}{c(1+z)D_M},
\end{equation}
while the smallest $k_{\perp}$ probed by the instrument is related to the antenna extent $d$, and is given by
\begin{equation}
k_{\perp min}=\frac{2\pi df_{HI}}{c(1+z)D_M}.
\end{equation}
The highest value of $k_{||}$ sampled by the instrument is related to the frequency resolution and is given by
\begin{equation}
k_{|| max}\approx  \eta_{max}\,\frac{2\pi H_of_{HI}E(z)}{c(1+z)^2}\,\,,\,\, \eta_{max}=\frac{1}{\Delta f}.
\end{equation}
The lowest value of $k_{||}$ sampled by the instrument in a snapshot observation is related to the frequency bandwidth and is given by
\begin{equation}
k_{|| min}\approx  \eta_{min}\,\frac{2\pi H_of_{HI}E(z)}{c(1+z)^2}\,\,,\,\, \eta_{min}=\frac{1}{\mathcal{B}}.
\end{equation}
\\

The final comment in this section is related to the implications of $lf$ invariance of the PSF to our understanding of the location and level of contamination in the instrumental $k$ space. Consider the foreground contamination in the neighborhood ($\delta l << 1$) of some image pixel $\mathcal{P}$ due to a point source at some $l_o$ removed from $\mathcal{P}$. Due to the $lf$ invariance, the transverse structure of contamination around $\mathcal{P}$ is identical to the line of sight contamination at $\mathcal{P}$. Since $\frac{\delta l}{l_o}=\frac{\delta f}{f_o}$, the energy in transverse fluctuations around $\mathcal{P}$ at wavenumber $k_{\perp \circ}$ will be the same as the energy in line of sight fluctuations at $\mathcal{P}$ at wavenumber $\eta_o=\frac{l_o}{f_o}\,k_{\perp \circ}$. Using Equation \ref{eqn:scales}, the $(k_{\perp},\,\,k_{||})$ pairs with identical contamination in them are defined by $k_{||}=k_{\perp}\,\,\frac{2\pi H_o E(z)}{c(1+z)}\,l_o$. For a uniformly weighted visibility distribution with complete coverage, there is significant energy only in the transverse mode $k_{\perp}=x_{max}\,f_o/c$. This gives us a corresponding $\eta_o=l_o/f_o\,k_{\perp}=x_{max}l_o/c$: a result we established in Section \ref{subsec:localization}.
\section{Effect of gridding and imaging algorithms}
\label{sec:gridding_algo}
While Section \ref{sec:fully_filled} described the nature of foreground contamination along the line of sight and transverse wavenumber axes, its results are a good description of the contamination in the special case of an ideal array, and hence lay the foundations needed to appreciate issues related to EoR detection with practical array geometries and data processing algorithms. Real arrays seldom have complete visibility coverage, and the Fourier relationships between visibility and image space, and between frequency and line of sight wavenumber space are, in practice, evaluated as discrete transforms. Computationally efficient transforms like the Fast Fourier Transform (FFT) are often used in Fourier synthesis arrays and almost always involve re-gridding the visibility data on a regular $uv$ grid. This section discusses the effects of such gridding and Fourier transform algorithms used for imaging on the nature of the 3 dimensional PSF and the resulting foreground contamination in $k$ space. The first generation EoR experiments propose to differentiate confusing foregrounds from the EoR signal by attempting to synthesize a confusion-free `EoR window' in the line of sight wavenumber dimension. We now describe the effect of gridding and imaging algorithms on the foreground contamination in $\eta$.\\

We have shown in Section \ref{sec:fully_filled} that contamination in $\eta$ space is substantially different in the two regions (see Figures \ref{fig:rel_sll} and \ref{fig:kspace}) $\mathcal{R}_1$ and $\mathcal{R}_2$, where $\mathcal{R}_1$ extends from $\eta =0$ to $\eta=x_{max}/c$, and $\mathcal{R}_2$ (or `EoR window') extends from $\eta=x_{max}/c$ to $\eta=1/\Delta f$. We have also shown that contamination in $\mathcal{R}_2$ is due to the sidelobes of the convolution kernel $\tilde{B}$ (see Equation \ref{eqn:final_relationship}) and the delta-function like peaks from sources at various directions are confined to $\mathcal{R}_1$. Consequently, if region $\mathcal{R}_2$ exists ($1/\Delta f > x_{max}/c$), as it should in any well-designed EoR instrument that aims to synthesize a confusion-free `EoR window', then PSF variation in frequency at any direction cosine ${\bf l}$ is oversampled by the instrument spectral resolution $\Delta f$. This implies that, in the ideal case, there will not be significant channel to channel variation in flux versus frequency at any sky pixel. However, processes like gridding may induce channel to channel `jitter' in the PSF and potentially contaminate $\mathcal{R}_2$. This stochastic component of foreground contamination arising from the gridding process may be termed `gridding contamination'. We now describe such stochastic effects of gridding in terms of channel to channel migration of baseline vectors from one $uv$ pixel to another. 

\subsection{Baseline migration}
\label{subsec:bline_migration}
\begin{figure}[h]
\epsscale{1.0}
\plotone{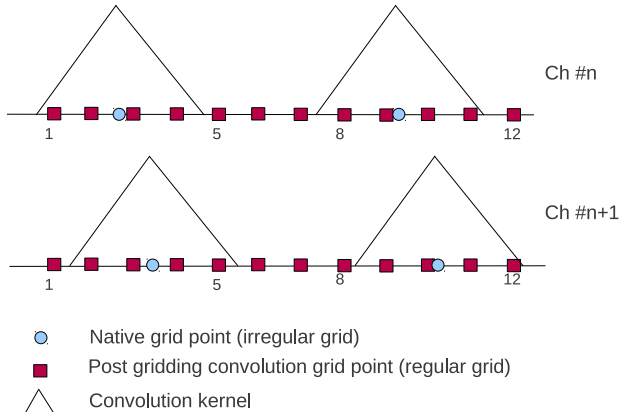}
\caption{Depiction of inter-channel baseline migration in one dimension ($u$). Red squares are grid points. Blue circles are baseline vector locations in wavelength units. The continuous line represents the post gridding convolution visibility distribution for a triangular convolution kernel. For simplicity, we have assumed that both depicted visibility measurements have equal weights. The upper panel shows the visibility axis for some frequency channel $n$, and the lower panel shows the visibility axis for the adjacent frequency channel $n+1$. The depicted visibility segment is shown to have a fairly sparse $u$ coverage for clarity. Notice how the baselines migrate on the visibility grid. While grid pixel number 1 and 8 go from being `filled' to being a `hole' between channel $n$ and $n+1$, visibility pixels 5 and 12 go from being a `hole' to being `filled'. Pixels like 2 and 10 continue to be filled but experience a `step' change in their weighted visibility value. \label{fig:pixel_migration}}
\end{figure}
Fourier synthesis arrays have traditionally formed image cubes in $lmf$ space by (i) computing the baselines (in wavelength units: ${\bf u}={\bf x}.f/c$) generated by a given antenna distribution at every frequency channel, (ii) weighting the visibilities to get desirable PSF characteristics, (iii) performing gridding convolution on the visibilities to re-grid them on a regular $uv$ grid at every frequency channel,  and (iv) using an efficient FFT algorithm to compute a `dirty' sky map at every frequency channel. Among these signal processing algorithms, gridding convolution has not been considered in the analytical treatment of Section \ref{sec:analytical}. We now describe the influence of gridding convolution on foreground contamination in $\eta$ space.\\

A baseline vector ${\bf x}$ (in meter units) will suffer a displacement of $\delta {\bf u}={\bf x}\delta f/c$ (in wavelength units) between adjacent frequency channels in the $uv$ plane. Here $\delta f$ is the inter-channel frequency spacing. We call this phenomenon `baseline migration'. This displacement results in `step' changes in gridded visibility values of pixels within the support of the convolving kernel. Many such `step' changes in visibilities due to the displacement of many baseline vectors results in a stochastic `jitter' in the frequency structure of the PSF. PSF `jitter' due to baseline migration may induce channel to channel fluctuations in flux versus frequency on the image plane, thereby significantly contaminating the `EoR window' $\mathcal{R}_2$.  This PSF `jitter' due to `baseline migration' is expected to have high contribution (i) from longer baseline vectors due to the relatively higher quantum of baseline displacement $\delta {\bf u}$, and (ii) from regions of low $uv$ density. In regions of higher $uv$ density a given $uv$ pixel, during gridding convolution, receives contributions from several neighboring baselines. The visibility contribution to a given $uv$ pixel then depends on the relative displacement of the neighboring baselines, rather than their actual displacement. On the other hand, consider the extreme case of very low $uv$ density as depicted in Figure \ref{fig:pixel_migration}. Here the actual inter-channel displacement of the baseline vectors induces relatively high `steps' in the post-gridding convolution visibilities. Natural weighting results in significantly lower weight assigned to regions with (i) longer baseline vectors and (ii) low $uv$ density and is expected to result in low PSF `jitter' in frequency. Uniform weighting, on the other hand, increases the weighting for longer baselines and is expected to result in substantially high PSF `jitter'. 
\subsection{The central $uv$ void}
\label{subsec:central_void}
The angular power spectrum of Extragalactic sources has been observed to have a power-law like form \citep{blake02} with substantially greater power on small angular scales; the universe is homogeneous and isotropic on large scales. Consequently, short baselines are not expected to have substantially greater response to Extragalactic continuum sources. However, the mean sky brightness will have a response at zero spacing ($u=0, v=0$). Energy in various spatial Fourier modes is seldom perfectly isolated and the zero spacing component will leak into short baselines. Consequently, channel to channel changes in the visibility distribution close to $u=0,v=0$ may result in `step' changes in flux versus frequency at image pixels thereby contaminating higher values of $\eta$ in $\mathcal{R}_2$. While Extragalactic sources are expected to have a smooth distribution on large angular scales, flux from the Galactic synchrotron emission is expected to follow a power law with significant spatial distribution power at large angular scales. Consequently short baselines close to $u=0,v=0$ will have significant response to our Galaxy and `baseline migrations' in this part of the $uv$ plane will create significant PSF `jitter' that may contaminate $\mathcal{R}_2$. \\

Fourier synthesis arrays usually have a central `$uv$ void' as the antenna elements have a minimum spacing due to constraints imposed by their own physical size. Antenna cross-talk considerations often result in array configurations with a central `$uv$ void' that is several times the antenna size. In such cases, the central $uv$ void spans several $uv$ pixels. We mentioned in Section \ref{subsec:bline_migration} that `baseline migrations' in regions of high $uv$ density do not contribute significantly to inter-channel PSF `jitter'. Though short spacing baselines lie in a densely populated part of the $uv$ plane, with increasing frequency, most baseline vectors are displaced radially away from any $uv$ grid point close to $u=0,v=0$, as opposed to other $uv$ grid points where baseline vectors also get displaced towards them. This asymmetric neighborhood near short spacing visibilities results in relatively higher inter-channel `step' changes in the visibilities on short baseline vectors. To alleviate this problem, the central $uv$ void may be chosen to be larger than that imposed by antenna cross-talk constraints.\\

\subsection{The Chirp Z transform}
\label{subsec:czt}
We have seen how channel to channel changes in the visibility weighting $W({\bf u},f)$ contaminates the line of sight wavenumber dimension. Here $W({\bf u},f)$ includes the effect of all visibility weighting and tapers used, and $W({\bf u},f)$ determines the PSF via the imaging transform. If the visibility distribution has complete coverage, maintaining invariance of $W({\bf u},f)$ with frequency is possible through application of a frequency independent visibility taper function. In such a case, $W({\bf u},f)$ is same at all frequencies and there is no frequency dependence in the PSF. In practice, there are many holes in the $uv$ grid and \citet{panos09} have suggested that a visibility mask may be used that rejects the $uv$ pixels that are not filled at all (or most) frequencies prior to image formation. At one end, we may exploit the complete instrument data and sensitivity at the cost of increased contamination in $\mathcal{R}_2$, and at the other end we may have a frequency independent PSF with some loss in information and sensitivity. However, there exists a `middle path' that assures that no data is rejected while restricting the foreground contamination in $\mathcal{R}_2$---image formation using the Chirp Z Transform, or CZT \citep{rabiner69}. We first present a brief description of CZT.\\

\begin{figure}[h]
\epsscale{1.0}
\plotone{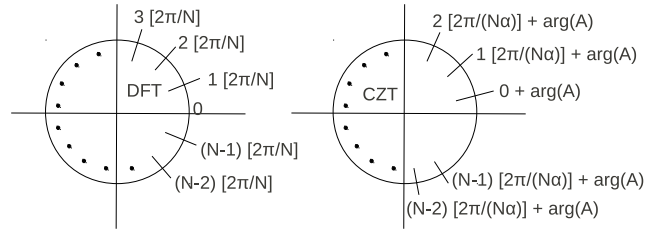}
\caption{Sketches showing points on the $z$ plane on which the generalized Chirp Z transform is computed for two cases: Case 1 (left panel) is for $A=1$ and $W=\exp\left(-i2\pi/N\right)$ and Case 2 (right panel) is for $|A|=1$ and $W=\exp\left(-i2\pi/(N\alpha)\right)$. In Case 1, the generalized CZT reduces to DFT, and for Case 2 the generalized CZT reduces to a special case of CZT that will be used in the proposed imaging algorithm of Section \ref{subsec:czt_algo}.\label{fig:loci}}
\end{figure}
The $Z$ transform of any finite discrete sequence $x_n,\,\,n=0,1,2....N-1$ is defined as
\begin{equation}
X(z)=\sum_{n=0}^{n=N-1}x_nz^{-n}.
\label{eqn:zt_def}
\end{equation}
$Z$ transform is a function of the complex variable $z$. When evaluated at discrete points along the unit circle defined by $z_k=\exp\left(i2\pi k/N\right)$, where $k=0,1,2....N-1$,  $Z$ Transform of the finite discrete sequence $x_n$ reduces to its Discrete Fourier Transform (DFT). The Chirp Z Transform or CZT is another special case of the $Z$ transform that is evaluated at discrete points on the locus given by $z_k=AW^{-k}$ where $A$ and $W$ are complex constants. Such a locus describes a spiral in the complex $z$ plane. Notice how for $A=1$ and $W=\exp\left(-i2\pi/N\right)$ the CZT again reduces to a DFT.\\

Consider a case where $|A|=1$ and $W=\exp\left(-i2\pi/(N\alpha)\right)$ where $\alpha$ is a real constant. The locus of points on the complex $z$ plane where the transform is evaluated now lie on a unit circle like in the case of a DFT. However, angular spacing between the points on the locus is now $2\pi/(n\alpha)$ as opposed to $2\pi/N$ in case of a DFT. The first point on the locus is also shifted by the argument of the complex constant $A$. The loci for the two cases are shown in Figure \ref{fig:loci}. Though $|A|=1$, $W=\exp\left(-i2\pi/(N\alpha)\right)$ is a special case of the generalized Chirp Z Transform, we refer to it hereinafter as Chirp Z Transform or CZT with a chirp factor $\alpha$. The only difference between DFT and CZT is the presence of the chirp factor $\alpha$ and an offset due to the presence of $A$. $\alpha$ simply `stretches' the locus on which the transform is evaluated and the CZT of a sequence is essentially a DFT of a `stretched' version of the sequence---the stretching factor being $\alpha$. We now describe how visibilities can be gridded in meter units and how a chirp factor can be used to compensate for the `stretching' of baseline vectors with frequency.\\ 

\subsection{A CZT based imaging algorithm}
\label{subsec:czt_algo}
As described in Section \ref{subsec:bline_migration}, the principal cause of channel to channel PSF variations in conventional gridding and imaging processes is baseline migration due to independent visibility gridding at every channel. However, the baseline vector generated by a given antenna pair moves across the visibility grid with change in frequency in a very predictable manner. The baselines in wavelength units at any frequency $f$ are merely the baselines in meter units `stretched' by a factor $\alpha$, where $\alpha=f/c$. In principle, we may grid the visibilities only once in meter units and use the CZT with chirp factor $\alpha$ to form the image at every frequency channel. Given an image extent, the grid size in meter units is chosen to avoid aliasing even at the highest frequency in the visibility cube. Gridding the visibilities only once in meter units eliminates the problem of non-linear baseline migrations and consequently ameliorates the problem of contamination in $\mathcal{R}_2$. In summary, we grid the baselines in meter units, and thereby do not let the baselines scale with frequency, and a given antenna pair will fall on exactly the same visibility grid point at all frequencies. We then compute the image on an $lm$ grid that is constant over frequency and use the chirp factor $\alpha$ in a CZT to compensate for an unchanging baseline. Such a CZT based imaging transform is given by
\begin{eqnarray}
S({\bf l}_{mn}) & = & \sum_{i=0}^{i=N-1}\sum_{j=0}^{j=N-1}V_1({\bf x}_{ij},f) \nonumber \\
                &   & \exp\left({-\frac{j2\pi}{N\alpha(f)}\,{\bf x}_{ij}{\bf l}_{mn}}\right),
\label{eqn:modified_fft}
\end{eqnarray} 
where $i,j$ are $uv$ (meter units) grid pixel indicies, $m,n$ are image grid pixel indicies, function $V_1({\bf x}_{ij},f)$ represents the visibilities at frequency $f$ gridded on to the common grid, and $\alpha(f)$ is the chirp factor given by $\alpha(f)=c/f$. The transform in Equation {\ref{eqn:modified_fft} is similar to a DFT and may be evaluated by the same efficient FFT algorithms with a minor modification to the exponential terms. The CZT image formation results in reduced contamination in $\mathcal{R}_2$ and, moreover, at a substantially reduced computational cost of gridding convolution because the gridding convolution at all frequency channels is done to a single common grid with little additional computational cost for imaging. 
\section{The MWA example}
\label{sec:mwa_example}
We have so far discussed the nature of foreground contamination in $\eta$ space for an ideal array with complete visibility coverage. We have also outlined some of the practical considerations applicable to realistic arrays and in the process proposed a new imaging technique using the Chirp Z Transform. In this section we demonstrate the relationships and results of Sections \ref{sec:analytical}, \ref{sec:fully_filled}, and \ref{sec:gridding_algo} using the case of snapshot imaging with MWA. MWA consists of 512 antenna elements called `tiles'. Each tile consists of a phased array of 16 bow-tie antennas. The tiles are distributed to form baselines up to about $1$ km. The instantaneous bandwidth is about $30$ MHz, and we assume a frequency resolution of about $40$ kHz. MWA will produce snapshot images every $8$ seconds. The instrumental response for this instrument configuration in the $k_{\perp}-k_{||}$ plane is shown in Figure \ref{fig:kspace}. Details of the MWA telescope design may be found in \citet{colin09}.

\subsection{Confusing sources and sidelobes}
\label{subsec:sonf_sources}
To estimate `blending confusion', we use the source counts given by \citet{subr02}. The threshold flux density $S_C$ above which we expect to find one source per beam of full width at half maximum $\theta_{arcmin}$ at frequency $f_{GHz}$ is given by
\begin{equation}
S_C = 40\,\theta_{arcmin}^{1.67}f^{-0.67}_{GHz} \,\,\, \mu\text{Jy}.
\end{equation}
$S_C$ is a confusion limit for a survey with excellent visibility coverage and all sources much stronger than $S_C$ can, in principle, be identified and subtracted. The stochastic response to sources weaker than $S_C$ contribute to sky pixel rms that is close to $S_C$. $S_C$ for MWA evaluates to $2.3$ mJy and is about $3$ orders of magnitude higher than the expected thermal noise in $8$-second snapshot images: MWA snapshot images will be confusion limited. For our simulations we assume successful peeling of point sources above $5S_C\approx10$ mJy. The expected residual rms confusion noise after successful subtraction of sources above $10$ mJy is about $3.6$ mJy. Due to the dominance of confusion over thermal noise, we will neglect the effects of thermal noise hereinafter. However, it must be noted that post foreground modeling and successful subtraction, the residual image that will be used for EoR power spectrum estimation is expected to be dominated, among other residual fluctuations, by thermal noise.\\ 

The threshold to which continuum sources may be identified and subtracted also depends on our ability to distinguish spillover flux due to sidelobes on the sky. We now roughly estimate the expected rms fluctuation at any image pixel due to the sidelobes on all the confusing sources in the sky--- a quantity we call `sidelobe confusion'. If we assume (i) the sidelobes of the PSF to be a zero mean random variable, and (ii) the confusion flux in each pixel to be a random variable that is uncorrelated with the sidelobe level, then the approximate rms sidelobe confusion can be estimated fairly easily: $C_S=\sum_i P_i S_i$ where $P_i$ is the sidelobe level at pixel $i$, and $S_i$ is the primary beam weighted true sky flux due to confusing sources at pixel $i$. Since $\langle P_i \rangle=0$, the variance in sidelobe confusion is $\langle C_S^2 \rangle = \sum_i \langle P_i^2 S_i^2 \rangle = \sum_i\langle P_i^2 \rangle\,\langle S_i^2 \rangle$. The far sidelobes of MWA are expected to have an rms variation of about $0.05\%$ giving $\langle P_i^2 \rangle=0.25$x$10^{-6}$, and the RMS confusion is about $3.6$ mJy giving $\langle S_i^2 \rangle = 13$ mJy$^2$. The sidelobe level assumed here is true for sidelobes far from the phase center at which `sidelobe confusion' is being computed. In reality sidelobe levels vary from sky pixel to pixel and a constant value has been assumed here for simplicity. Far sidelobe level has been chosen since far sidelobes result in foreground contamination at higher values of $\eta$ where we expect to detect the EoR with minimal contamination, and this calculation will present the worst case. While $P_i$ is assumed to be independent of $i$, $S_i$ is assumed to be the confusion noise weighted by the primary beam gain, and the summation is performed using the expected primary beam of the MWA primary antenna element (Full width to first null of $60^{\circ}$) to evaluate `sidelobe confusion'. The resulting `sidelobe confusion' from the entire sky is $\langle C_S \rangle \approx 1.9$ mJy. The similarity in the values of `blending confusion' and `sidelobe confusion' is an interesting consequence of the similarity between the number of resolution elements in the image (weighted by the primary beam) and inverse of the sidelobe level of the synthesized beam. This implies that after subtracting bright point sources from MWA snapshot images, the resulting residual image has statistical fluctuations with roughly equal contributions from the stochastic response to the Extragalactic sources and the sidelobe levels.\\

The EoR brightness fluctuations are expected to have an rms of about $25$ mK on scales comparable to the MWA $5$ arcmin resolution. This corresponds to an rms flux of about $36\,\mu$Jy. Hence, apart from confusing foreground sources, `sidelobe confusion' can confuse any attempt to detect the EoR signal through a simple measurement of image plane variance. That said, flux from confusing sources and PSF sidelobes have structure in the frequency domain which can be exploited to detect the EoR fluctuations despite a high rms fluctuation in the image plane. While spectral structure of confusion Extragalactic sources follow relatively simple power laws, spectral structure of instrumental sidelobe confusion is more complex as described in Sections \ref{sec:analytical}, \ref{sec:fully_filled}, and \ref{sec:gridding_algo}. We now quantitatively estimate the foreground contamination in frequency space (and eventually in $\eta$ space) for the case of snapshot imaging with MWA.
\subsection{A snapshot observation}
\label{subsec:snapshot_obs}
This subsection describes a simulation to estimate the foreground contamination in $\eta$ space for the case of snapshot imaging of Extragalactic point sources using MWA. The simulation assumes a frequency invariant sky model and assumes that all sources brighter than $10$ mJy have been successfully subtracted. The sky model is then interpolated into an $lm$ grid with grid size at least 2 times finer than the MWA instrument resolution. The grid extent is covers the interval $[-0.5\,\,0.5]$ in direction cosines that corresponds to the MWA primary beam full width at first null, which is $60^{\circ}$ at $150$ MHz. Gridding convolution of visibilities is performed using a triangular kernel like the one shown in Figure \ref{fig:pixel_migration}. A simple kernel was chosen purely to illustrate the nature of contamination in $\eta$ space. At every frequency channel, three separate PSFs are generated using natural weighting, uniform weighting, and uniform weighting with a Gaussian taper (hereinafter referred to simply as Gaussian weighting). The simulated sky in $lm$ is then multiplied with the PSF at zenith and the result is summed. This accumulated value at every frequency channel gives the dirty image flux density at the zenith pixel. This zenith pixel flux versus frequency is used to evaluate the spillover of foreground contamination  by Fourier transforming to $\eta$ domain. The simulation was performed for three cases each involving a different gridding and imaging algorithm; we discuss these below.\\

Case 1 employed independent gridding convolution at every frequency channel, and the dirty image was formed using an efficient FFT algorithm. Figure \ref{fig:spill_nug_normal} shows the flux density versus frequency and contamination power in $\eta$ for this case.
\begin{figure*}[h]
\epsscale{1.0}
\centering
\includegraphics[width=6in]{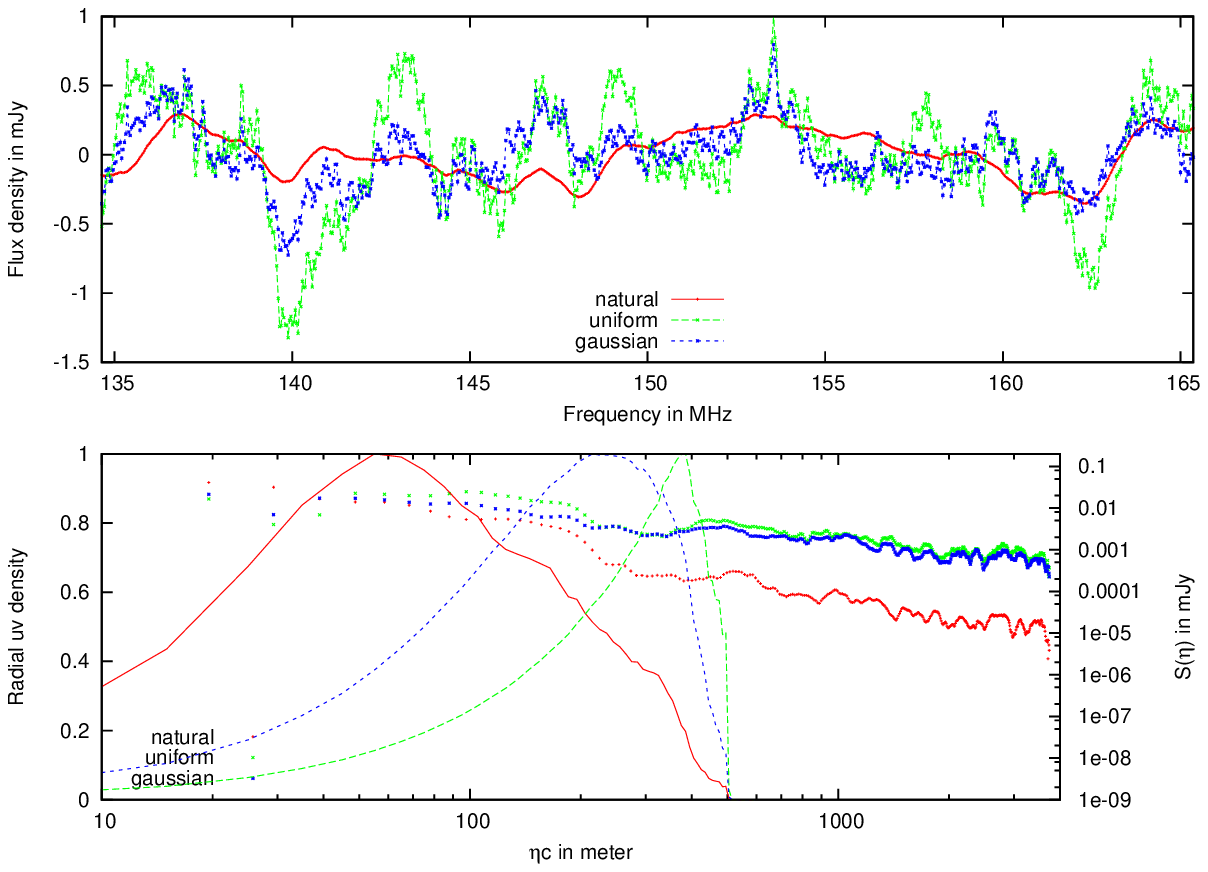}
\caption{Foreground contamination along the line of sight dimension in the zenith pixel for the case of snapshot imaging with MWA. In generating these representations of foreground contamination, we populate sources in the sky out to $|{\bf l}|_{max}=0.5$. The zenith pixel flux is plotted against frequency (upper panel) and $\eta$ (lower panel). Notice the jitter in the upper panel for the case of uniform weighting and Gaussian weighting as opposed to the case of natural weighting. This jitter results in significant `gridding contamination' at higher values of $\eta$ as seen in the lower panel. Also shown in the lower panel is radial $uv$ density as a function of $|{\bf x}||{\bf l}|_{max}$ for the case of natural, uniform and Gaussian weighting. The boundary between $\mathcal{R}_1$ and $\mathcal{R}_2$ lies at $\eta c = |{\bf x}|_{max}|{\bf l}|_{max} = 500$ meter.\label{fig:spill_nug_normal}}
\end{figure*}
\begin{figure*}[h]
\epsscale{1.0}
\centering
\includegraphics[width=6in]{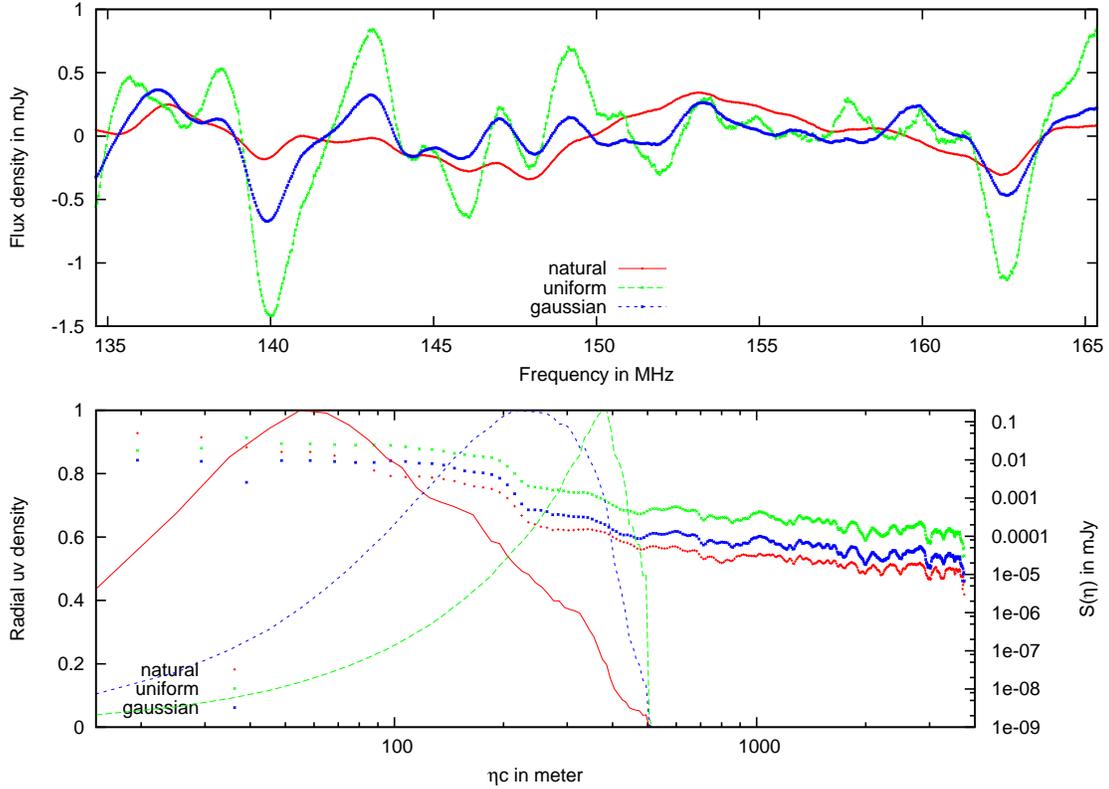}
\caption{Foreground contamination along the line of sight dimension in the zenith pixel for the case of snapshot imaging with MWA using the Chirp Z Transform (CZT). Notice how the jitter evident in Figure \ref{fig:spill_nug_normal} has significantly reduced in the zenith pixel flux density versus frequency plot (upper panel). Consequently, foreground contamination at higher values of $\eta$ (lower plot) is significantly lower as compared to that in Figure \ref{fig:spill_nug_normal}. A finite amount of undesirable `jitter' still persists since a different set of baselines vectors were used in the gridding and imaging process at every frequency channel so as to keep the final image resolution constant at all frequencies. \label{fig:spill_nug_czt}}
\end{figure*} 
\begin{figure*}[h]
\epsscale{1.0}
\centering
\includegraphics[width=6in]{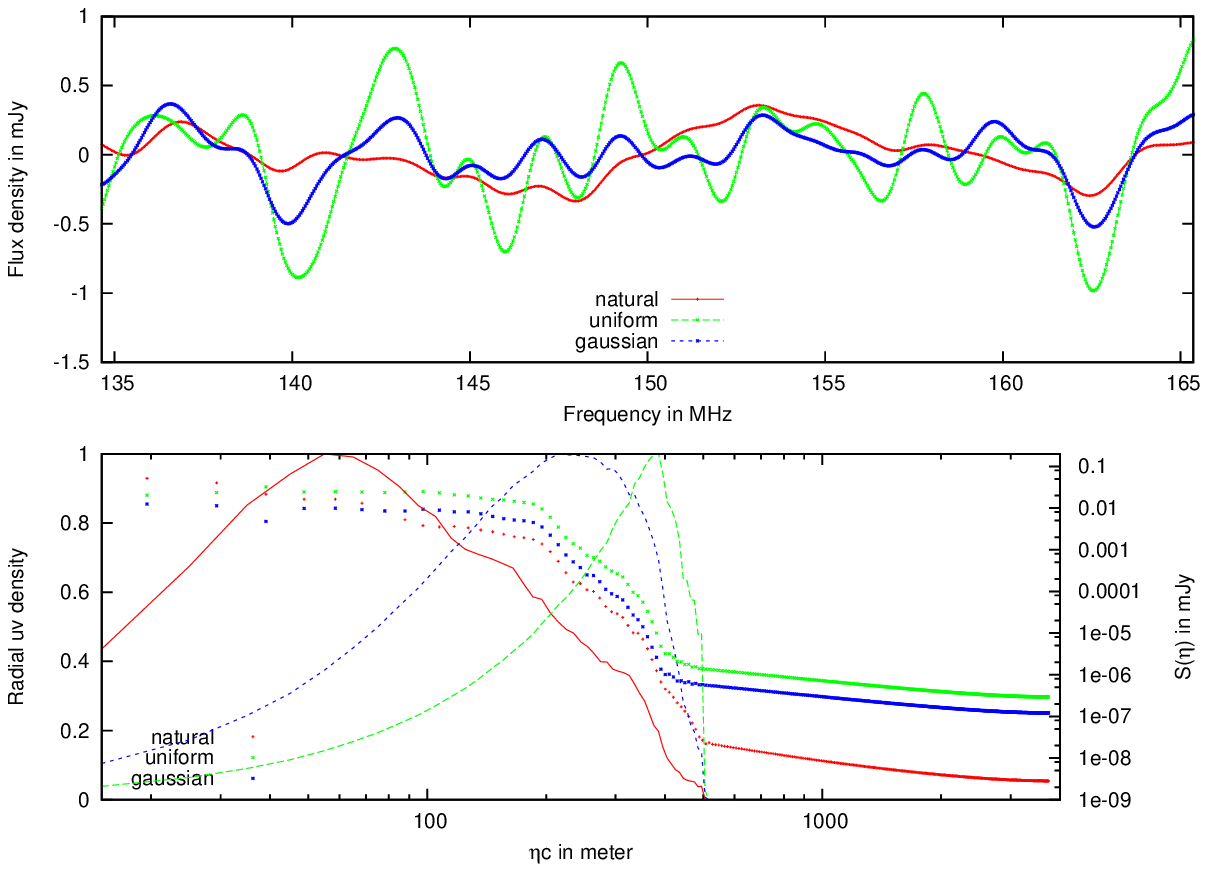}
\caption{Foreground contamination along the line of sight dimension in the zenith pixel for the case of snapshot imaging using the Chirp Z Transform (CZT) with frequency dependent image resolution. Notice how there is negligible `jitter' in the flux density versus frequency (upper panel). The 'gridding contamination' at higher values of $\eta$ (lower panel) in $\mathcal{R}_2$ is now wholly owing to `PSF contamination'.\label{fig:spill_nug_czt_nodiff}}
\end{figure*} 
We have shown earlier that for $\eta < x_{max}/c$ in $\mathcal{R}_1$, the exact structure of spillover depends on the nature of relative weightings given to visibilities. Weighting here includes the effect of $uv$ plane distribution of baselines, user defined weighting and any tapers applied. For the case of uniformly weighted visibilities with complete $uv$ coverage, the $\eta$ spillover from a point source is expected to be localized since the convolution in Equation \ref{eqn:final_relationship} has a significant contribution only due to the discontinuity in the weighting function at $|{\bf x}|=x_{max}$. In case of MWA snapshot visibilities, the weighting function is not `smooth' for $|{\bf x}|<x_{max}$ and the convolution in Equation \ref{eqn:final_relationship} has a significant contribution from all $|{\bf x}|<x_{max}$. Though this results in a distributed spillover, a significant part of the spillover energy is still restricted to $\mathcal{R}_1$, and $\mathcal{R}_2$ will, in the absence of baseline migrations, be contaminated only due to the sidelobes of the convolving kernel from Equation \ref{eqn:final_relationship} as discussed earlier. This is the `PSF contamination' that is expected to persist in $\mathcal{R}_2$. Any additional channel to channel variations in the PSF will contaminate $\mathcal{R}_2$ beyond the `PSF contamination'. Such channel to channel variations are evident in the upper panel of Figure \ref{fig:spill_nug_normal} in the form of `jitter', especially for the cases of uniform weighting and Gaussian weighting. This `jitter' was found to be due to inter-channel baseline migrations as discussed in Section \ref{sec:gridding_algo}. To quantitatively estimate the amount of `gridding contamination' due to baseline migration, the Blackman-Nuttall window with very low sidelobes was used as a frequency window function $B(f)$. As described in Section \ref{subsec:sidelobes}, such a windowing function will significantly suppress `PSF contamination' in $\mathcal{R}_2$ ($\eta c > 500$ meter), and the dominant residual contamination in $\mathcal{R}_2$ will be due to the inter-channel `jitter' arising from the gridding process.\\

It is interesting to note that baseline migrations have induced relatively higher jitter for the cases of uniform weighting and Gaussian weighting as opposed to natural weighting. As discussed in Section \ref{subsec:bline_migration} this is due to two reasons. First, for an inter-channel frequency spacing of $\delta f$, a baseline vector ${\bf x}$ (in meter units) suffers an inter-channel migration on the $uv$ plane of length ${\bf x}\delta f/c$ (in wavelength units). Second, and more importantly, a relatively higher amount of inter-channel `jitter' due to baseline migrations arises from regions of lower $uv$ density (longer baseline vectors) compared to regions of higher $uv$ density (smaller baseline vectors). Since uniform weighting, and in this case Gaussian weighting, place higher emphasis on longer baseline vectors as compared to natural weighting, the amount of `jitter' in Figure \ref{fig:spill_nug_normal} goes in decreasing order of magnitude from uniform weighting to Gaussian weighting to natural weighting. Earth rotation synthesis and drift scan-synthesis will increase the $uv$ density and is expected to ameliorate the `jitter'.\\

Case 2 involves image formation with the CZT algorithm with a similar image resolution at every frequency. As described in Section \ref{subsec:czt_algo}, image formation with the CZT algorithm is expected to lower `gridding contamination' in $\mathcal{R}_2$ arising from inter-channel jitter due to `baseline migrations'. Figure \ref{fig:spill_nug_czt} shows the line of sight contamination for the case of image formation using the CZT algorithm. The `jitter' in flux density versus frequency (upper plot) has significantly reduced due to the absence of baseline migrations inherent to independent gridding at every frequency channel---an expected outcome of gridding the visibilities in meter units. Similar image resolution at every frequency channel was achieved by scaling the Gaussian taper to have the same form in wavelength units, and restricting the longest baseline vector that was used in the gridding and imaging process to the same value (in wavelength units) at every frequency channel. This resulted in inter-channel `steps' in the weighting function at the far end of the $uv$ plane that resulted in a small but significant amount of `jitter' and hence a contamination in $\mathcal{R}_2$. Nevertheless, the contamination in $\mathcal{R}_2$ for this case is reduced by a factor of $1.7$, $6.4$, and $21.7$ for natural weighting, uniform weighting, and Gaussian weighting respectively. Relative contamination for natural weighting has not improved significantly as regions in the $uv$ plane that contribute most to `jitter' are down-weighted by natural weighting.\\

Case 3 involves image formation with the CZT algorithm with a frequency dependent image resolution. In this case, the set of baselines that are used in the gridding and imaging routine are identical at all frequencies. It may be noted that despite a frequency dependent image resolution, the CZT algorithm computes the sky flux density on the same $lm$ grid at all frequencies. The results of this simulation are shown in Figure \ref{fig:spill_nug_czt_nodiff}. The contamination in $\mathcal{R}_2$ has dropped considerably owing to the elimination of channel to channel `jitter' in the PSF by this method. The extremely low sidelobes of the Blackman-Nuttall window and the absence of `baseline migration' in the CZT based imaging algorithm we have proposed herein result in extremely low levels of contamination in $\mathcal{R}_2$. Quantitatively, the contamination in $\mathcal{R}_2$ for this simulation as compared to Case $1$ has reduced by more than three orders of magnitude for all types of weighting schemes. \\

It is important to note that the amount of inter-channel `jitter' is largely dependent on the visibility distribution and in particular on the filling fraction in different parts of the $uv$ plane. Therefore, the relative merits of the three different weighting schemes discussed will depend on the array configuration and observing strategy. It is however expected that the contamination in $\mathcal{R}_2$ will be orders of magnitude lower if gridding and imaging are done using the proposed CZT algorithm along with a good frequency window. The simulations and results in this section are not exhaustive, and are included as a means to appreciate the various factors that influence the frequency dependence of the PSF and illustrate the potential reduction in foreground contamination.
\section{Conclusions and future work}
\label{sec:conclusions}
Redshifted $21$ cm tomography using Fourier synthesis arrays has emerged as a promising tool for EoR studies. Mitigating the confusion effects of Galactic and Extragalactic foregrounds is a major challenge to such EoR experiments. Through this work we have furthered understanding of contamination arising from Extragalactic continuum foreground and instrumental effects using a common framework. In particular, we have derived analytical expressions for contamination along frequency, or equivalently line of sight (LOS) dimension, which relates the frequency-axis contamination in the image cube to the visibility weighting in Fourier space. In doing so, we have cast the problem in Fourier space where the telescope measurements indeed lie, thus enabling us to draw conclusions that are directly applicable to instrument design and data processing.\\

We identified two major sources of foreground contamination---`PSF contamination' and 'gridding contamination', and described their structure in LOS wavenumber space. The `PSF contamination' due to any point source is localized in LOS wavenumber space around $\eta=x_{max}l/c$ where $l$ is the direction cosine of the source with respect to the imaged pixel, and $x_{max}$ is the maximum baseline length in meters. Consequently, when imaging with continuum confusion that is offset in direction cosine upto a maximum of $l_{max}$ from an imaged pixel, most of the `PSF contamination' will be confined to a regime $\mathcal{R}_1$ defined by $\eta \in [0\,\,\,\,x_{max}l_{max}/c]$. It is useful to note that there does exist a regime $\mathcal{R}_2$ defined by $\eta\in[x_{max}l_{max}/c\,\,\,\,\,1/\Delta f]$, where $\Delta f$ is the frequency resolution, where finite `PSF contamination' persists. This contamination in $\mathcal{R}_2$ may be suppressed by three orders of magnitude by judicious choice of a window function in frequency. For this reason, $\mathcal{R}_2$ is an `EoR window' where we may focus our efforts to detect EoR. \\

We have shown that gridding and imaging routines result in additional `gridding contamination' in LOS wavenumber space. In particular, we have shown that independent gridding at every frequency channel may lead to a stochastic inter-channel `jitter' in the PSF that potentially contaminates the `EoR window' $\mathcal{R}_2$. To ameliorate problems associated with inter-channel `jitter' due to gridding, we proposed an alternative gridding and imaging algorithm based on the Chirp Z Transform (CZT). In this algorithm the visibilities are gridded in meter units rather than wavelength units, thereby eliminating `gridding contamination'. The CZT compensates for the stretching of baselines (in wavelength units) with frequency by introducing a `chirp' term---a scaling factor by which the frequency of the Fourier sinusoids are multiplied. We finally demonstrated the localization `PSF contamination' and the elimination of `gridding contamination' in the CZT based imaging algorithm using simulations of imaging with MWA.\\

Our ongoing and future work is directed towards using the concepts and ideas in this paper in an all sky simulation of MWA images that incorporates different gridding convolution kernels, Earth rotation and drift-scan synthesis, effects of array non-coplanarity, and the Galactic synchrotron emission. We are currently developing specific simulation pipelines that will accomplish the all sky simulations with the CZT based imaging algorithm. Such simulations will quantitatively estimate not only the merit of the CZT based imaging algorithm for EoR detection, but also the expected contamination from foreground continuum in future MWA data products.

\end{document}